\documentclass[11pt]{article}

\usepackage{subfigure}
\usepackage{amssymb,amsmath}
\usepackage[dvips]{graphicx}
\usepackage{epsfig}
\usepackage{cite}

\def\laq{~\raise 0.4ex\hbox{$<$}\kern -0.8em\lower 0.62
ex\hbox{$\sim$}~}
\def\gaq{~\raise 0.4ex\hbox{$>$}\kern -0.7em\lower 0.62
ex\hbox{$\sim$}~}
\def\be{\begin{equation}}
\def\ee{\end{equation}}
\def\bea{\begin{eqnarray}}
\def\eea{\end{eqnarray}}
\newcommand{\nn}{\nonumber}
\newcommand{\de}{\partial}

\def \ra {\rightarrow}

\def \d {\delta}

\def \b {\beta}
\def \a {\alpha}
\def \b {\beta}

\def \r {\rho}
\def \vp {\varphi}
\def \om {\omega}

\def \5 {{}^{(5)}}

\def \dag {\dagger}

\def \dag {\dagger}
\def \Hef {H_{\rm eff}}
\def \Ref {R_{\rm eff}}
\def \aef {a_{\rm eff}}

\begin{document}

\begin{titlepage}

\begin{flushright}
BA-TH/11-642\\
\end{flushright}

\vspace*{1.5 cm}

\begin{center}

\Huge
{Cosmological backreaction of heavy string states}

\vspace{1cm}

\large{G. De Risi}

\normalsize
\vspace{.2in}
{\sl Dipartimento di Fisica, Universit\`a degli Studi di Bari, \\
and} \\
{\sl Istituto Nazionale di Fisica Nucleare, Sezione di Bari\\
Via G. Amendola 173, 70126 Bari, Italy}

\vspace*{1.5cm}

\begin{abstract}

We propose a mechanism to have a smooth transition from a pre-Big Bang phase to a standard cosmological
phase. Such transition is driven by gravitational production of heavy massive string states that
backreact on the geometry to stop the growth of the curvature. Close to the string scale, particle creation
can become effective because the string phase space compensate the exponential suppression of the particle production.
Numerical solutions for the evolution of the Universe with this source are presented.

\end{abstract}
\end{center}

\end{titlepage}
\newpage

\section{Introduction}
\label{Intro}

The standard cosmological model, even in its inflationary extension, is plagued by an ``initial condition problem''
\cite{Borde:1993xh}, since it must emerge from an initial singularity. One of the most successful proposal to overcome
this problem was the introduction of the so-called pre-Big Bang scenario \cite{Gasperini:1992em} (for a review,
see \cite{Gasperini:2002bn}), developed in the context of string theory, in which the Universe is supposed to emerge
from a string vacuum (so the initial geometry is flat), perturbed by quantum fluctuations of the dilaton (other
proposals were put forward in different contexts, such as brane cosmology and Loop Quantum Gravity, see for example
\cite{Khoury:2001wf,Steinhardt:2001st,Mukherji:2002ft,DeRisi:2007dn,Ashtekar:2006rx,Ashtekar:2006uz,DeRisi:2007gp}).
This is achieved by implementing purely stringy symmetries on the string cosmology equations. Those symmetries
suggest the existence of a phase of growing curvature. However, it is not clear, a priori how to connect smoothly
the pre-big bang and the post-big bang phases, since, at the classical level, the two phases are still disconnected
by a singularity. Many ideas were proposed to overcome this problem, all involving further assumptions
on the dynamics of the Universe, namely higher order correction (either loop or $\a'$ contribution) and
negative energy density contribution to obtain the desired ``repulsive gravity'' (an incomplete list would include
\cite{Gasperini:1996in,Gasperini:1996fu,Brustein:1997cv,Maggiore:1997vw,Cartier:1999vk,DeRisi:2001ed}).

In this paper we will propose a mechanism to obtain a regular evolution of the curvature
that rely on production of heavy massive string states by gravitational backreaction. It is well known
\cite{Birrel} that an evolving Universe will produce particles because of the squeezing of the ground
state. It has been speculated in the literature that this effect, though being negligible, could
actually have some physical consequences \cite{Lawrence:1995ct,Gubser:2003vk,Chung:1998bt}. The idea
is that the exponential suppression of the energy density of particles created during the evolution of the Universe
is compensated by the exponential growing of the multiplicity of states above the string scale (the Hagedorn spectrum),
thus giving a sensible contribution to the total energy density. Of course, dealing with gravitationally induced
particle production brings one to deal with a sort of semiclassical realization of quantum gravity,
which is not under control, and in fact the authors themselves of the mentioned papers admit some of the
ideas are rather speculative. Here we want to go one step forward with speculations, and assume that
the gravitationally produced energy backreacts locally on the geometry of the Universe. We will see that
even a rough estimate such as the one we are going to show here is enough to stop the growing of the
Ricci curvature and induce a bell-shaped behaviour. Let us stress that all the analysis is performed under
the assumtpion that the universe stays in a low-energy (low curvature) regime in which supergravity is
a valid approximation, and the only ingredient taken from stringy physics is the behaviour of the high energy spectrum
that influences the phase space of low curvature universe. In fact, we are going to treat string massive modes as
scalars, assuming however that their multiplicity is controlled by the Hagedorn spectrum.
The validity of this assumption is verified by checking that the curvature found vith the modified
cosmological equations do not exceed the string scale (which has been set to unity in the numerical
evaluationsI.

We are aware that some of the passages we will show, though physically well-motivated, can be challenged
on a formal basis. Nevertheless, since our goal was only to check if a backreaction mechanism could provide for
a graceful exit without invoking any further assumptions and higher order modifications of the string cosmology
equations, we just skipped over formal complexities to have an estimate of the order of magnitude of the effect.
We plan to come back on all the mathematical issues (among other things we will discuss below) in a forthcoming paper.

The paper is organized as follows: Section \ref{QFT} is devoted to some formal developments that will lead us to
write the enregy density (and pressure) produced by gravitationally excited string states in term of a
time dependent Bogoliubov coefficient. In section \ref{energy} we will use this expression to find, under certain
assumption, a sound expression for the energy density, to be put into the string cosmology equations whcih are solved
numerically in section \ref{cosmology}. Finally, in section \ref{conclusions} we will comment on
our results and the (rather strong) approximations assumed in obtaining them, and on how to further develop
the present study in several directions.

\section{QFT approach to gravitational backreaction}
\label{QFT}

Let us then consider a massive scalar field living in a curved space. The action is
\be
S = \frac{1}{2} \int d^4 x \sqrt{\left| g \right|} \left(g^{\mu \nu} \de_\mu \phi \de_\nu \phi
- m^2 \phi^2 \right).
\label{Action}
\ee
Let us assume that the geometry of the Universe is a conformally flat FRW; so, in the conformal time gauge,
the metric is
\be
ds^2 = a^2(\eta) \left(d\eta^2 - {\bf dx}^2 \right),
\label{confmetric}
\ee
and the action can be written as
\be
S = \frac{1}{2} \int d\eta d^3 x~a^2(\eta) \left(\phi'^2 - (\nabla \phi)^2 - m^2a^2(\eta)\phi^2 \right),
\label{confAction}
\ee
where $(\nabla \phi)^2$ stands for $\sum_i (\de_i \phi)^2$. Now we introduce the canonical
field $\vp (\eta, {\bf x}) = a(\eta) \phi (\eta, {\bf x})$, so that the action can be rewritten as
\be
S = \frac{1}{2} \int d\eta d^3 x\left[ \eta^{\mu \nu} \de_\mu \vp \de_\nu \vp -
\left( m^2 a^2(\eta) -\frac{a''(\eta)}{a(\eta)}\right) \vp^2 \right].
\label{canAction}
\ee
This action can be interpreted as describing a scalar field living on a flat space-time, with a potential
which controls its gravitational interactions. From the action (\ref{canAction}) we can evaluate
the (canonical) energy momentum tensor.
\bea
T_\mu^{~~\nu} &=& \frac{\de \mathcal{L}}{\de(\de_\nu \vp)}\de_\mu \vp-\mathcal{L}\d_\mu^{~~\nu} \nn \\
&=& \de_\mu \vp \de^\nu \vp - \frac{1}{2} \left[ (\de_\mu \vp)^2 - \left( m^2 a^2  - \frac{a''}{a}\right)
\vp^2 \right] \d_\mu^{~~\nu}.
\label{emtensor}
\eea
The diagonal elements of this tensor are the energy density and the pressure. The field conjugate to $\vp$ is
\be
\pi(\eta, {\bf x}) = \frac{\de \mathcal{L}}{\de \vp'},
\label{conjfield}
\ee
so the Hamiltonian density, to be identified with the energy density, and the pressure along the $i^{th}$ direction
are, according to (\ref{emtensor})
\bea
\r_m(\eta,{\bf x}) &=& \frac{1}{2} \left[\pi^2 + (\nabla \vp)^2 + V(\eta) \vp^2 \right], \nn \\
p_{m,i}(\eta,{\bf x}) &=& \frac{1}{2} \left[\pi^2 + (\nabla \vp)^2 - V(\eta) \vp^2 \right],
\label{enpressdens1}
\eea
where, of course, $V(\eta) = m^2 a^2(\eta) -\frac{a''(\eta)}{a(\eta)}$

Now we expand the fields in Fourier modes and, following conventional quantization procedure, we
promote ($\eta$-dependent) Fourier coefficients to operators:
\bea
\vp(\eta,{\bf x})&=&\frac{1}{(2\pi)^{3/2}}\int d^3k~ \vp_k(\eta) e^{i{\bf k}\cdot{\bf x}}, \nn \\
\pi(\eta,{\bf x})&=&\frac{1}{(2\pi)^{3/2}}\int d^3k~ \pi_k(\eta) e^{i{\bf k}\cdot{\bf x}},
\label{modeexp}
\eea
where $\pi_k(\eta)=\vp'_k(\eta)$ and $\vp_k(\eta)$ satisfies the (operatorial) evolution equation:
\be
\vp''_k(\eta) + \om^2(\eta)\vp_k(\eta)=0,
\label{modeeq}
\ee
with $\om_k(\eta)=\sqrt{k^2+V(\eta)}$. Moreover, since the field $\vp$ is real, and thus Hermitean, the modes must
satisfy the relation
\bea
\vp^\dag_k(\eta)&=&\vp_{-k}(\eta), \nn \\
\pi^\dag_k(\eta)&=&\pi_{-k}(\eta).
\label{moderel}
\eea
We can then write down the energy density and pressure in terms of these modes.
By defining the vector
\be
\zeta_k(\eta)= \left( \begin{array}{c} \vp_k(\eta) \\ \pi_k(\eta) \\ \end{array} \right),
\label{vecfipi}
\ee
the densities (\ref{enpressdens1}) can be rewritten as
\bea
\r_m(\eta,{\bf x}) &=& \frac{1}{2} \int \frac{d^3k~d^3p}{(2\pi)^3}\zeta_k^\dag (\eta)
~P_{k,p}(\eta)~\zeta_p(\eta) e^{i({\bf k}-{\bf p})\cdot{\bf x}}, \nn \\
p_{m,i}(\eta,{\bf x}) &=& \frac{1}{2} \int \frac{d^3k~d^3p}{(2\pi)^3}\zeta_k^\dag (\eta)
~Q^{(i)}_{k,p}(\eta)~\zeta_p(\eta) e^{i({\bf k}-{\bf p})\cdot{\bf x}},
\label{enpressdens3}
\eea
with
\bea
P_{k,p}(\eta) &=& \left( \begin{array}{cc} {\bf k}\cdot {\bf p} + V(\eta) & 0\\0 & 1  \\ \end{array} \right), \nn \\
Q^{(i)}_{k,p}(\eta) &=& \left( \begin{array}{cc} {\bf k}\cdot {\bf p} - V(\eta) & 0\\0 & 1  \\ \end{array} \right).
\label{P}
\eea
This expression is useful, because we know \cite{Bozza:2003pr,Bozza:2002fp} that the evolution for
$\zeta_k(\eta)$ is given by:
\be
\zeta_k(\eta)=U_k(\eta,\eta_0)\zeta_k(\eta_0),
\label{zetaevoleq}
\ee
where the propagator can be written as
\be
U_k(\eta,\eta_0) = \left( \begin{array}{cc} A_k(\eta,\eta_0) & B_k(\eta,\eta_0), \\
C_k(\eta,\eta_0) & D_k(\eta,\eta_0) \\ \end{array}\right),
\label{Umatrx}
\ee
and the four coefficients of the matrix are
\bea
A_k(\eta,\eta_0) &=& i \left[g_k(\eta_0) f_k^*(\eta) - g_k^*(\eta_0) f_k(\eta) \right], \nn \\
B_k(\eta,\eta_0) &=& i \left[f_k(\eta) f_k^*(\eta_0) - f_k^*(\eta) f_k(\eta_0) \right], \nn \\
C_k(\eta,\eta_0) &=& i \left[g_k(\eta_0) g_k^*(\eta) - g_k^*(\eta_0) g_k(\eta) \right], \nn \\
D_k(\eta,\eta_0) &=& i \left[g_k(\eta) f_k^*(\eta_0) - g_k^*(\eta) f_k(\eta_0) \right].
\label{ABCD}
\eea
Here $f_k(\eta)$  is solution of the canonical evolution equation (\ref{modeeq}) for modes of the field $\vp$
and $g_k(\eta)$ is solution of the mode equation (which we have not reported) for the conjugate momentum
$\pi$ (so, of course, the relation $g_k(\eta) = f'_k(\eta)$ holds), while $\eta_0$ is a suitable time on which
we will define initial condition. We will assume that $\eta_0$ is far enough in the past so that
it is possible to consider the space to be Minkowski. Using eq. (\ref{zetaevoleq}) we can thus write:
\bea
\r_m(\eta,{\bf x}) = \frac{1}{2} \int \frac{d^3k~d^3p}{(2\pi)^3}\zeta_k^\dagger(\eta_0)~S_{k,p}(\eta,\eta_0)~
\zeta_p(\eta_0) e^{i({\bf k}-{\bf p})\cdot{\bf x}}, \nn \\
p_{m,i}(\eta,{\bf x}) = \frac{1}{2} \int \frac{d^3k~d^3p}{(2\pi)^3}\zeta_k^\dagger(\eta_0)~T^{(i)}_{k,p}(\eta,\eta_0)~
\zeta_p(\eta_0) e^{i({\bf k}-{\bf p})\cdot{\bf x}},
\label{enpressdens4}
\eea
where,
\bea
S_{k,p}(\eta,\eta_0)&=&U_k^\dagger(\eta,\eta_0)P_{k,p}U_k(\eta,\eta_0), \nn \\
T^{(i)}_{k,p}(\eta,\eta_0)&=&U_k^\dagger(\eta,\eta_0)Q^{(i)}_{k,p}U_k(\eta,\eta_0).
\label{S}
\eea
Now we want to express this in terms of creation-annihilation operators. Notice that on an actual
Minkowski space, there are no ambiguities in building a Fock space; so we define:
\be
\zeta_k(\eta_0)=\left(\begin{array}{c}\frac{1}{\sqrt{2\om_k}} \\ -i\sqrt{\frac{\om_k}{2}} \\ \end{array}\right) a_k
+ \left(\begin{array}{c}\frac{1}{\sqrt{2\om_k}} \\ i\sqrt{\frac{\om_k}{2}} \\ \end{array}\right) a^\dag_{-k},
\label{creanform}
\ee
with $a_k$ and $a^\dag_k$ satisfying the usual canonical commutation rules and $\om_k \equiv \om_k(\eta_0)$.
Inserting this expression in (\ref{enpressdens4}) we get
\bea
\r_m &=& \frac{1}{2} \int \frac{d^3k~d^3p}{(2\pi)^3}\left[ S^{(1)}_{k,p}a^\dag_k a_p +
S^{(2)}_{k,p}a^\dag_k a^\dag_{-p} + S^{(3)}_{k,p}a_{-k} a_p +
S^{(4)}_{k,p}a_{-k} a^\dag_{-p}\right]e^{i({\bf k}-{\bf p})\cdot{\bf x}}, \nn \\
p_{m,i} &=& \frac{1}{2} \int \frac{d^3k~d^3p}{(2\pi)^3}\left[ T^{(i,1)}_{k,p}a^\dag_k a_p +
T^{(i,2)}_{k,p}a^\dag_k a^\dag_{-p} + T^{(i,3)}_{k,p}a_{-k} a_p + T^{(i,4)}_{k,p}a_{-k} a^\dag_{-p}\right]
e^{i({\bf k}-{\bf p})\cdot{\bf x}}, \nn \\
\label{enpressdens5}
\eea
where the $S^{(l)}$ and $T^{(i,l)}$ are calculated from the bilinear product of the matrices
$S_{k,p}$ and $T^{(i)}_{k,p}$ respectively, with the vectors defined in
(\ref{creanform}). In taking the VEV of (\ref{enpressdens5}), only the term with the correct order survives. Then the
$\delta$ function allows us to integrate over one of the two (equal) momenta. Moreover, in order to handle
with this expression in a cosmological context (as we are going to do), we assume that particles are created
homogeneously throughout the space, so that the integral in (\ref{enpressdens5}) depends only on the square
momentum $k^2$ of the excitation. We thus get
\bea
\langle 0|\r_m(\eta) |0\rangle = 2 \pi \int_0^{+\infty} dk~k^2 S^{(4)}_{k,k}(\eta,\eta_0), \nn \\
\langle 0|p_{m,i}(\eta) |0\rangle = 2 \pi \int_0^{+\infty} dk~k^2 T^{(i,4)}_{k,k}(\eta,\eta_0).
\label{enpressdens6}
\eea
The coefficients $S^{(4)}$ and $T^{(i,4)}$ are easily calculated. After a little algebra we get:
\bea
S^{(4)}_{k,k}(\eta,\eta_0) &=& \frac{1}{2 \om_k} \left[ \left(k^2 + V(\eta)\right) \left|A_k(\eta,\eta_0)\right|^2
+ \left|C_k(\eta,\eta_0)\right|^2 \right] \nn \\
&& + \frac{\om_k}{2} \left[ \left(k^2 + V(\eta)\right) \left|B_k(\eta,\eta_0)\right|^2 +
\left|D_k(\eta,\eta_0)\right|^2 \right], \nn \\
T^{(i,4)}_{k,k}(\eta,\eta_0) &=& \frac{1}{2 \om_k} \left[ \left(k^2 - V(\eta)\right) \left|A_k(\eta,\eta_0)\right|^2
+ \left|C_k(\eta,\eta_0)\right|^2 \right] \nn \\
&& + \frac{\om_k}{2} \left[ \left(k^2 - V(\eta)\right) \left|B_k(\eta,\eta_0)\right|^2 +
\left|D_k(\eta,\eta_0)\right|^2 \right].
\label{S4andT4}
\eea
This expression can be further specified by noting that, if the space-time is flat in $\eta = \eta_0$,
solutions for the evolution equation (\ref{modeeq}) can be written as:
\bea
f_k(\eta \ra \eta_0) &=& \frac{1}{\sqrt{2 \om_k}} e^{-i\om_k(\eta - \eta_0)}, \nn \\
g_k(\eta \ra \eta_0) &=& -i\sqrt{\frac{\om_k}{2}} e^{-i\om_k(\eta - \eta_0)},
\label{flatfunct}
\eea
with $\om_k = \sqrt{k^2+m^2}$ ($m$ is the mass of the excitation). Furthermore, we get a better insight
on the expressions of energy density and pressure by writing the mode functions $f_k$
and $g_k$ in terms of time-dependent Bogoliubov operators \cite{Audretsch:1979uv,Gubser:2003vk}:
\bea
f_k(\eta)&=&\frac{\a_k(\eta)}{\sqrt{2\om_k(\eta)}}\exp\left( -i\int_{\eta_0}^\eta dx~\om_k(x) \right) +
\frac{\b_k(\eta)}{\sqrt{2\om_k(\eta)}}\exp\left( i\int_{\eta_0}^\eta dx~\om_k(x) \right), \nn \\
g_k(\eta)&=&-i \sqrt{\frac{\om_k(\eta)}{2}}\a_k(\eta)\exp\left( -i\int_{\eta_0}^\eta dx~\om_k(x) \right) +
 i\sqrt{\frac{\om_k(\eta)}{2}}\b_k(\eta)\exp\left( i\int_{\eta_0}^\eta dx~\om_k(x) \right),\nn \\
\label{fgtoab}
\eea
with $\a_k$ and $\b_k$ satisfying the condition $|\a_k|^2-|\b_k|^2=1$, which follows from the Wronskian condition on
$f_k$ and $g_k$. Substituting this in (\ref{S4andT4}), and plugging everything into the expressions for
the energy density and pressure (\ref{enpressdens6}) we get:
\bea
\r_m(\eta) &=& 2 \pi \int_0^{+\infty} dk~k^2 \left( 1+2|\b_k(\eta)|^2 \right)\om_k(\eta), \label{bendens} \\
p_m(\eta) &=& 2 \pi \int_0^{+\infty} dk~\frac{k^4}{\om_k(\eta)} \left( 1+2|\b_k(\eta)|^2 \right). \label{bpress}
\eea
In the next section, we will use this expressions to obtain the energy density as a function
of the geometry of the on-shell universe

\section{Energy density of the gravitationally excited string states}
\label{Energy}

In the previous section, we obtained a formal expression for the energy density and pressure of massive scalar modes
during the evolution of the universe. At this stage we can already notice that, since (\ref{bpress}) is suppressed
by a factor $m$ with respect to (\ref{bendens}), the pressure will be negligible with respect to the energy density
for large mass (non-relativistic) modes. Therefore, from now on we will assume that the pressure is exactly zero,
leaving a more detailed treatments for future works.

Asymptotically, the Bogoliubov coefficient $\b_k(\eta)$ can be interpreted as the number
density of particles produced via gravitational interaction from an initial vacuum state. Let us stress that, by
setting $\b_k(\eta) = 0$, which is true for $\eta \ra \eta_0$, expression (\ref{bendens}) reduces just to the
unrenormalized zero point energy of the field. Thus we are led to consider a properly renormalized form
for (\ref{bendens}) as the energy density of massive excitations of the field under consideration, generated by
gravitational interaction. We choose to regularize the energy density by subtracting the time-dependent zero-point
energy of the oscillations, so to get
\be
\r_m(\eta) = 4 \pi \int_0^{+\infty} dk ~k^2 |\b_k(\eta)|^2 \om_k(\eta).
\label{finalendens}
\ee
Such expression for the energy density has the key feature of vanishing at $\eta = \eta_0$ (as it is expected
since we want a vacuum solution to start with), and in general whenever the space-time is flat. In addition, it has
the expected form of a sum over energies of each mode\footnote{Note that $\om_k(\eta)$ is not exactly the energy
of the single mode, since it contains also the ``pump field'' term}, following the interpretation of $\b$ as the number
density of gravitationally produced particles, with the spectrum modified by gravitational interaction.

We now need to evaluate $\b_k(\eta)$, which is in general a very difficult task.
If we assume that the evolution of the Universe is ``slow'' enough, so that we are in an adiabatic
regime \cite{Birrel}, we can use for $\b_k(\eta)$ the approximate expression \cite{Chung:1998bt}
\be
\b_k(\eta) = \int d\eta \frac{\om'_k(\eta)}{2\om_k(\eta)}
\exp \left( -2i \int_{\eta_0}^\eta d \eta'\om_k(\eta') \right).
\label{beta_appr}
\ee
The integral can be evaluated, to the same level of approximation, with the steepest descendent method
\cite{Gubser:2003vk,Chung:1998bt} giving:
\be
|\b_k(\eta)|^2 = \exp \left[ -4 \frac{\om_k^2(\eta)}{V(\eta)\sqrt{\frac{6\Hef^2(\eta) - \Ref(\eta)}{6 m^2}}} \right],
\label{beta_final}
\ee
where $\Hef$ and $\Ref$ are respectively the Hubble parameter and the scalar curvature, calculated starting from an
effective scale factor
\be
\aef(\eta) = \sqrt{a^2(\eta) - \frac{a''(\eta)}{m^2 a(\eta)}}.
\label{a_eff}
\ee
With this expression for $\b_k$, the integral in (\ref{finalendens}) converges and can be evaluated exactly. After
some technical manipulations we get:
\be
\r_m(\eta) = -\frac{\pi^2}{8}V^2(\eta) \sqrt{\frac{6\Hef^2(\eta) - \Ref(\eta)}{6 m^2}}
H^{(1)}_1 \left( i\frac{2m}{\sqrt{\Hef^2(\eta) - \frac{\Ref(\eta)}{6}}} \right)
\exp \left( -\frac{2m}{\sqrt{\Hef^2(\eta) - \frac{\Ref(\eta)}{6}}} \right),
\label{int_en_dens}
\ee
where $H^{(1)}_1$ is the Hankel function of the first kind. For heavy massive states we can approximate
the Hankel function with its large argument expansion \cite{Gradshteyn}, thus getting:
\be
\r_m(\eta) = \frac{\pi^{3/2}}{8} V^2(\eta) \left( \frac{6\Hef^2(\eta) - \Ref(\eta)}{6 m^2} \right)^{3/4}
\exp \left( -\frac{4m}{\sqrt{\Hef^2(\eta) - \frac{\Ref(\eta)}{6}}} \right).
\label{large_en_dens}
\ee

As expected, the contribution for each massive mode is exponentially suppressed at low (sub-Planckian) energies.
Incidentally, being at sub-Planckian energies justifies our quantum field theory approach to production of
string modes (which, of course, have masses above that limit), as we stressed in the introduction.
Now, as we already mentioned above, the multiplicity of string modes is exponentially growing \cite{Gubser:2003vk}
$N \propto \exp(m/T_H)$ ($T_H$ being the Hagedorn temperature), and this can possibly compensate the exponential
suppression of each produced mode. In fact, the total energy density generated by all the stringy massive modes is:
\be
\r_{tot}(\eta) = \sum_{m \geq m_s} \r_m(\eta) \exp \left( \frac{m}{T_H}\right).
\label{sum_en_dens}
\ee
Again, this expression can be estimated by noting that $\aef(\eta) \simeq a(\eta)$ for heavy enough
states, and approximating the series as an integral, substituting $m \simeq T_H x$, with $x > 0$. We
get:
\bea
\r_{tot}(\eta) &\simeq& \frac{\pi^3}{8}T_H^{5/2}a^4 \left( H^2 - \frac{R}{6} \right)^{3/4}
\int_0^{+\infty}dx~x^{5/2}\exp\left[-x \left( \frac{T_H}{\sqrt{H^2 - \frac{R}{6}}} - 1 \right) \right] \nn \\
&\simeq& T_H^4 a^4 \left( \frac{H^2 - \frac{R}{6}}{T_H^2} \right)^{5/2}.
\label{tot_en_dens}
\eea
Since the energy density is a scalar, turning to cosmic time is straightforward. We finally get:
\be
\r_{tot}(t) \simeq T_H^4 a^4(t) \left( \frac{H^2(t) - \frac{R(t)}{6}}{T_H^2} \right)^{5/2}.
\label{fin_en_dens}
\ee
In the next section, we will use this expression as a source in string cosmolgy equations.

\section{Cosmology with the backreacting energy density}
\label{cosmology}

\begin{figure}[h]
\subfigure[]{\includegraphics[height=4.8cm]{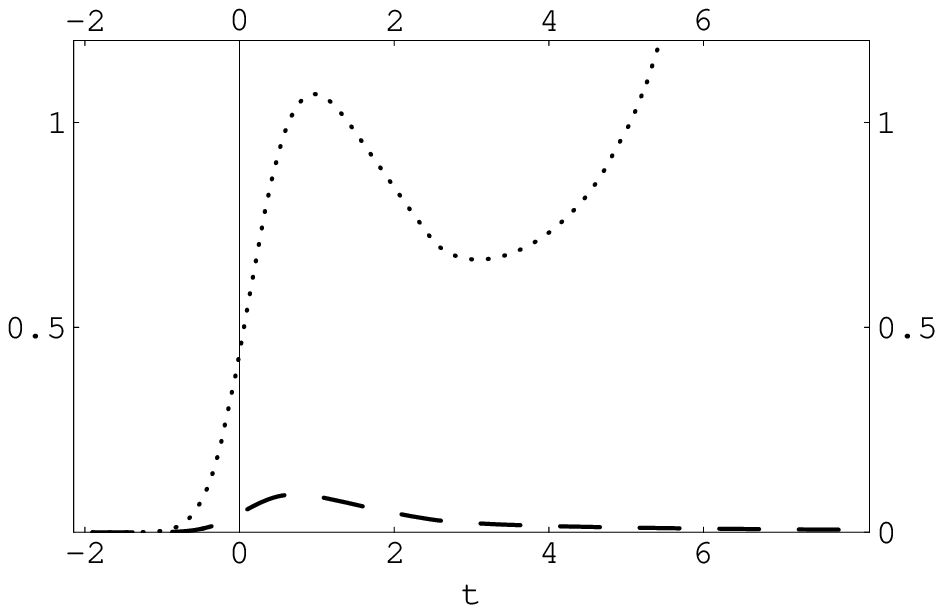}}
\hspace*{0.2cm}
\subfigure[]{\includegraphics[height=4.8cm]{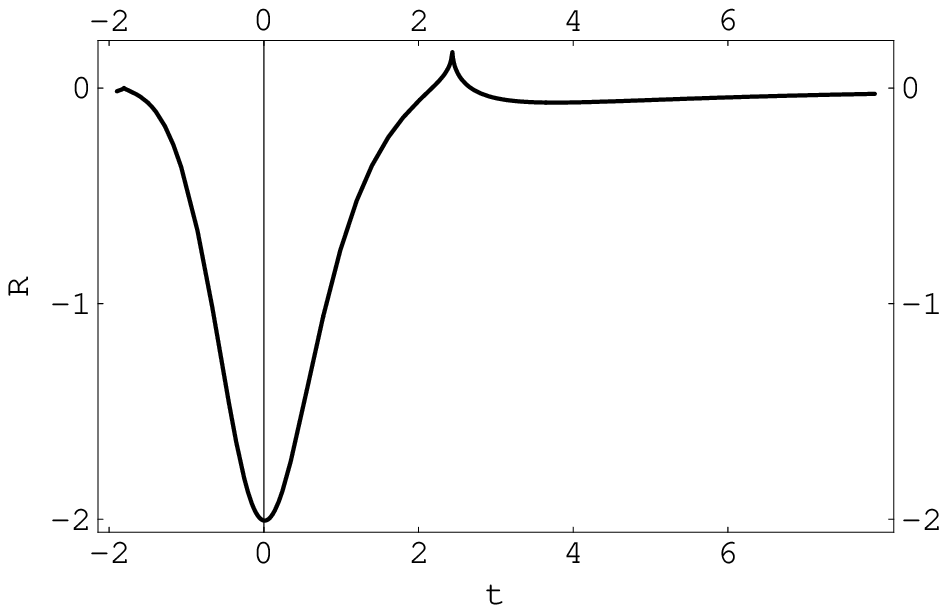}}
\subfigure[]{\includegraphics[height=4.8cm]{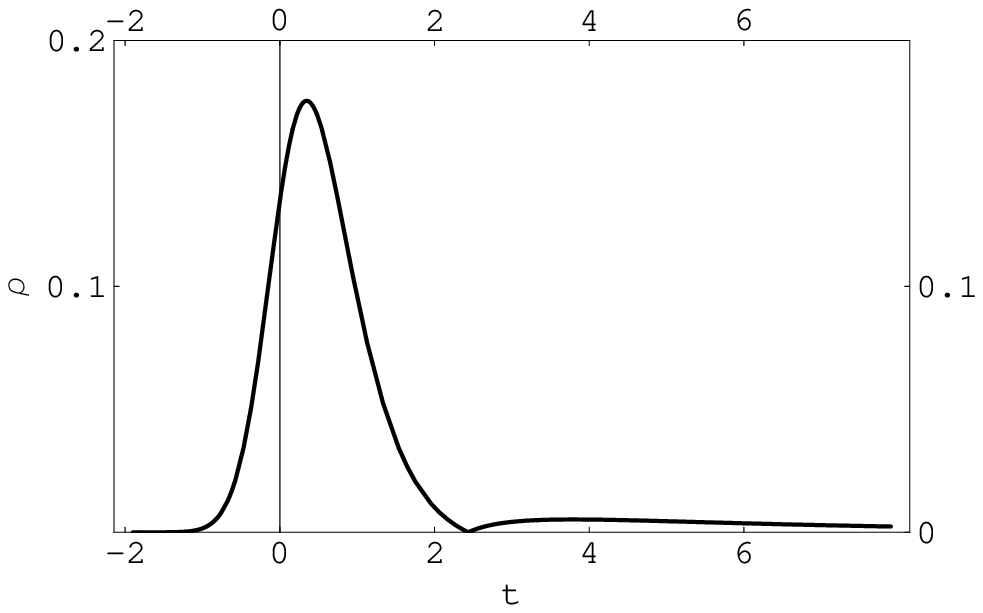}}
\hspace*{0.2cm}
\subfigure[]{\includegraphics[height=4.8cm]{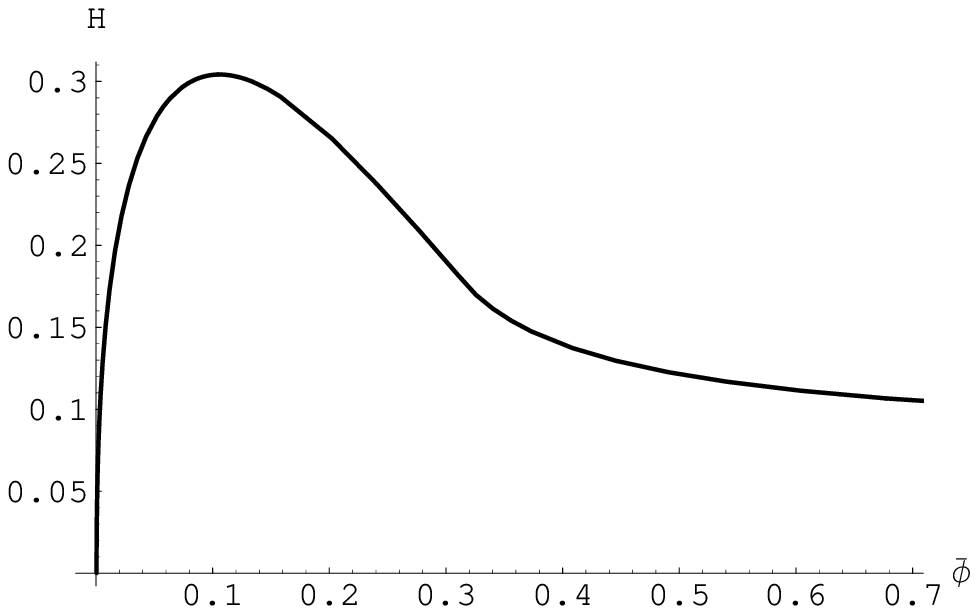}}
\caption{Plots of the dilaton energy density $\dot{\phi}^2$ (dotted line) and the squared Hubble parameter $H^2$
(dashed line) (a), the Ricci curvature $R$ (b), the energy density of the generated massive string states $\rho$ (c),
and the parametric plot of the ``Shifted dilaton'' energy density versus the Hubble parameter (d).
The Hagedorn temperature has been set to $1$.}
\label{Pictures}
\end{figure}

In the last section we have found an anaylitic expression for the energy density massive string states
excited by the gravitational interaction with an evolving universe.
Now we assume that this energy density backreacts on the geometry of the Universe to favor a smooth transition
between a phase of superinflationary accelerated expansion with growing curvature to a (standard) phase of decelerated
expansion and decreasing curvature. To see this, we plug the expression (\ref{fin_en_dens}) into
the pre-big bang equations. We assume that the extra dimension are compactified down to the Plank scale, and
there is a mechanism for the stabilization of the modula. The string cosmology
equations then read \cite{Gasperini:2002bn}
\bea
\dot{\phi}^2 - 6 H \dot{\phi} + 6 H^2 &=& \frac{2}{T_H^2} e^{\phi}\r_{tot},
\label{Einst_eq1} \\
2\ddot{\phi} + 6 H \dot{\phi} - \dot{\phi}^2 - 6 \dot{H} - 12 H^2 &=& 0.
\label{Einst_eq2}
\eea
(we have identified the string mass with $T_H$) where $\r_{tot}$ is given by (\ref{fin_en_dens})
and we have assumed that matter is minimally coupled to the metric of the string frame, without any direct
dilaton coupling, see, e.g. \cite{Gasperini:2001mr} (we will comment more on this later). Of course,
these equations cannot be solved analytically, so we must turn to numeric. Plots for the interesting physical
quantities are shown in Fig. \ref{Pictures}.
Numerical integration has been carried out by fixing a dummy initial timescale so that the
maximum value of the curvature occurs at $t = 0$, and by setting suitable
(small, in units of string mass) initial condition for the Hubble parameter and the dilaton energy density.
As we can see from panel (b), the curvature reaches a maximum and decreases, evolving towards a phase of
De Sitter evolution (i.e. with our notations, a constant curvature regime). It is possible to check,
in addition, that all the approximation requirements are met. On the other hand, the dilaton energy density,
depicted in panel (a) with the squared Hubble parameter is not under control and blows up, after reaching a
first local maximum at the same time age of the curvature (see also plot (d)). This is not surprising,
since we have not considered the coupling of the generated massive states with the dilaton and the
(possible) corresponding dilaton stabilization. Motivated by the successful transition to a phase of decelerated
expansion we found in our approach, we seek to extend the present analysis to the dilatonic sector in a
forthcoming paper, so to have a full-fledged graceful exit solution.

\section{Conclusions and Outlook}
\label{conclusions}

Particle creation generated by gravitational backreaction seems to provide a cogent
mechanism to stop superinflationary accelerated expansion and to drive the Universe to a (standard)
phase of decelerated expansion. Calculations we have presented in this paper,
though being speculative in some of their aspects, show that, as the curvature grows, there is the
possibility that gravitational energy excites heavy string states, thus producing some sort of
``friction'' effect. This can actually slow down and eventually stop the growth of the curvature and
prevent the occurrence of the singularity. This can happen before the curvature reaches the string scale, ao
before the universe enters in a regime in which quantum field theory  and supergravity are no longer valid.

However, the analysis we have performed needs to be further developed in many directions:
First of all, some of the mathematical derivations need to be posed on a more steady ground (for example,
our derivation rely on the assumption that the metric is conformally flat, so that
formal results valid in a flat space can be straightforwardly extended to our curved model).
In addition, as we already noted, the particle creation mechanism has to include a direct coupling to
the dilaton. In fact, even though the geometry is regular everywhere in the toy model we have presented, the
dilaton still blows up, indicating a breakdown of the validity of our approximations. Nevertheless,
since the dilaton itself is a string mode, it is expected to couple with massive states, so
high energetic dilaton modes will somehow excite heavy modes as well. This will probably provide a mechanism
to stabilize the dilaton after the transition.

 Moreover, to actually enter into a phase of standard evolution, one should consider some sort of
``reheating process''. The idea is that string states generated via gravitational backreaction decays into
radiation and matter, to start the standard phase. Another issue worth to be pursued is to have a complete
dynamical system analysis. In fact, our calculations confirm the known result that late-time
evolution of the Universe falls toward a de Sitter phase. It is not clear, however, if the de Sitter curvature scale
is controlled by the string scale, or can be influenced by backreaction. In the second case, maybe one can
trigger the mechanism (in addition to other contributions coming from a correct approach to modula and
extra-dimension stabilization) to lower the de Sitter scale towards some realistic value. In this way
we could possibly use the same tools that cure the pathological behaviour of the early Universe also
to address problems of our late time Universe.

\section*{Aknowledgments}
It is a pleasure to thank M. Gasperini, C. Ungarelli and G. Veneziano for discussions and comments on the manuscript.
We also wish to thank the ``Angelo Della Riccia'' foundation for financial support and the Theory division
at CERN, Geneva for warm hospitality at the initial stages of this work.

\end{document}